\documentclass[useAMS,usenatbib]{mnras}
\usepackage{amssymb}     \usepackage{amsmath}    \usepackage{amsfonts}
\usepackage{xspace}    \usepackage{longtable}   \usepackage{rotating} 
\usepackage{lscape}
\usepackage{times}
\title[Radio AGN contribution to the radio--FIR correlation]
  {Determining the radio AGN contribution to the radio--FIR correlation using the black hole fundamental plane relation}
\author[O.\ I.\ Wong et al.]
  {O. Ivy Wong,$^{1,2}$\thanks{E-mail:ivy.wong@uwa.edu.au} M.~J.~Koss,$^{3}$ K.~Schawinski,$^{3}$ A.~D.~Kapi\'{n}ska,$^{2,1}$ I.~Lamperti,$^{3}$ K.~Oh,$^{3}$
\newauthor C. Ricci$^{4}$ and S.~Berney$^{3}$ \\
  $^1$ International Centre for Radio Astronomy Research, The University of Western Australia M468, 35 Stirling Highway, Crawley, WA 6009, Australia\\
  $^2$ ARC Centre of Excellence for All-Sky Astrophysics (CAASTRO)\\
  $^3$ Institute for Astronomy, ETH Zürich, Wolfgang-Pauli-Strasse 27, 8093 Z\"urich, Switzerland\\
  $^4$ Pontificia Universidad Catolica de Chile, Instituto de Astrofsica, Casilla 306, Santiago 22, Chile; EMBIGGEN Anillo, Concepcion, Chile\\
}

\date{Released 2015 Xxxxx XX}

\pagerange{\pageref{firstpage}--\pageref{lastpage}} \pubyear{2016}

\def\LaTeX{L\kern-.36em\raise.3ex\hbox{a}\kern-.15em
    T\kern-.1667em\lower.7ex\hbox{E}\kern-.125emX}

\begin{document}

\label{firstpage}

\maketitle

\begin{abstract}

We investigate the 1.4 GHz radio properties of 92 nearby ($z<0.05$) ultra hard
 X-ray selected Active Galactic Nuclei (AGN) from the Swift Burst Alert Telescope (BAT)
 sample. Through the ultra hard X-ray selection we minimise the biases against obscured 
or Compton-thick AGN as well as confusion with emission derived from star formation
that typically affect AGN samples selected from the UV, optical and infrared wavelengths. 
We find that all the objects in  our sample of nearby, ultra-hard X-ray selected AGN 
are radio quiet; 83\% of the objects are classed as high-excitation galaxies (HEGs) and 
17\% as low-excitation galaxies (LEGs).
While these low-$z$ BAT sources follow the radio--far-infrared correlation in a 
similar fashion to star forming galaxies,  our analysis finds that there is still significant AGN 
contribution in the observed radio emission from these radio quiet AGN. In fact, the 
majority of our BAT sample occupy the same X-ray--radio fundamental plane as have
been observed in other samples, which include radio loud AGN ---evidence that the observed
radio emission (albeit weak) is connected to the AGN accretion mechanism, rather than star 
formation.
%We  hypothesise that the observed radio properties of our sample can be explained by an 
%accretion disk corona origin. 

\end{abstract}


\section{Introduction}

Much progress has been made towards  understanding the co-evolution of 
central supermassive black holes (SMBHs) and their host galaxies  in both theoretical simulations
\citep[e.g.\ ][]{sesana14,hopkins15} and observations \citep[e.g.\ ][]{kormendy13,heckman14}.
Recently, it is becoming apparent that SMBHs (like their host galaxies) 
 follow two distinct evolutionary modes: (1) one that exhibits
 a high excitation and high efficiency accretion; and (2) another with a low excitation and 
low efficiency accretion \citep{hopkins14,gatti14,menci14}. 
The accreting SMBHs at the centre of galaxies are known as Active Galactic Nuclei (AGN).

While much overlap exists between samples, the AGN populations
identified through soft X-rays, ultraviolet (UV )/optical, mid-infrared (MIR) and radio
 observations result in largely distinct samples.  This is largely due to the 
different origin of the observed emission, dust obscuration and orientation of the
AGN with respect to the observer's line-of-sight \citep[e.g.\ ][]{antonucci93,urry95}.
Typically, soft X-ray- and UV/optical- selected samples of AGN are biased against
obscured AGN \citep{mushotzky04}, while strong MIR emission can be confused with emission
from star formation \citep{stern05,hickox09}.  Emission at the shorter wavelengths 
(soft X-ray and UV/optical) originate from regions closer to the central SMBHs (corona 
and/or the accretion disk), while the dominant source of MIR emission is the dusty 
torus \citep[e.g.\ ][]{manners02}.

\citet{griffith10} studied a sample of radio-, X-ray, and MIR-selected AGN identified
within the COSMOS survey. Though the soft X-ray selection is biased against obscured AGN,
the MIR and radio selection is less biased against obscured AGN.  From this sample of 
mostly unobscured AGN, the optical brightness appears to scale with that observed in the 
X-ray and MIR wavelengths---suggestive that the  X-ray- and mid-IR-selected AGNs have
 high Eddington ratios, similar to those of optically-selected AGN. On the other hand,
the radio brightness of radio-selected AGN are not typically correlated with their optical 
brightness. Many low power radio-selected AGN do not exhibit narrow emission lines in 
the optical wevelengths \citep{laing94,jackson97}; can lack the IR emission from a dusty 
torus \citep{whysong04,ogle06}; and may not show accretion-related X-ray emission
\citep{Hardcastle06, Evans06}. Hence, it is generally thought that radio-selected
AGN represents a SMBH growth mode that involves accretion mechanism(s) with a 
lower and wider range of Eddington ratios \citep{hickox09,griffith10,schawinski15}.

%Current studies have found  AGN luminosities in the optical wavelengths to be
%correlated and scale with those in the X-ray and mid-infrared wavelengths \citep{griffith10}.
% Such results suggest  that the  X-ray- and mid-IR-selected AGNs have similarly high Eddington ratios.
%However, the radio luminosity densities of radio-selected AGNs do not correlate
%with the AGN properties in the optical wavelengths and is hence thought to represent a SMBH 
%growth mode that involves an accretion mechanism with a lower and wider range of Eddington ratios
%  \citep{hickox09,griffith10}.

On the other hand, the radio and X-ray emission from luminous radio-selected AGN have 
long been understood to be closely-linked due to the  assumed coupling between the accretion disk 
and the occasional presence of radio outflows ---as described by the black hole 
``fundamental plane'' correlation 
\citep{merloni03,falcke04} observed between the X-ray luminosities ($L_{\rm{X}}$), 
radio luminosities ($L_{\rm{R}}$) and the mass of the SMBH.  With few exceptions 
\citep{coriat11,burlon13,panessa13,dong14}, this correlation holds for radio loud AGN on 
scales of  parsecs \citep[e.g.\ ][]{merloni03} to kiloparsecs \citep{panessa15} as well as 
accreting stellar mass black holes (as in the case of X-ray binaries).  
{\em{Therefore, how does the black hole ``fundamental plane'' correlation fit into the current
paradigm of bimodality in AGN samples?}}
 
To investigate the physical mechanisms that are driving the observed emission at X-ray and radio 
wavelengths, we study a sample of nearby ultra hard X-ray  selected Seyfert AGN from the Swift 
BAT survey ($14-195$~keV) compiled by \citet{koss11a} (hereafter known as the K11 sample). While a 
hard X-ray selection from emission observed at 2--10~keV is sufficient at revealing  hidden 
or obscured narrow-line AGN \citep[e.g.\ ][]{ueda07}, an ultra hard X-ray selection at energies greater than
15~keV will reveal even Compton-thick sources (with obscuring column densities $>10^{24}$~cm$^{-2}$)
and provide a more complete census of the AGN population. At the ultra hard energy bands $>15$~keV,
 we also avoid any possible emission contribution from star formation.

The radio observations in this paper {\bf{are}} obtained from two large surveys at 1.4~GHz using the 
Very Large Array located in New Mexico:  1) the Faint Images of the Radio Sky at Twenty-centimeters 
\citep[FIRST; ][]{becker95,white97} survey; and 2) the NRAO VLA Sky Survey \citep[NVSS; ][]{condon98}.
 The 1.4~GHz observations will be analysed in combination with the X-ray and far-infrared 
properties of the host galaxy in order to shed light on the radio properties (and origin) of 
 a more complete sample of low-redshift AGN.  However, 
we do not include beamed AGN such as blazars in our sample because we are interested in studying 
the co-evolution between the host galaxy and its accreting central SMBH.

We describe our ultra hard X-ray selected sample of nearby AGN from the Swift BAT survey in
Section 2.1.  Section 2.2 describes the complementary 1.4 GHz radio observations used in this paper.  
Section 3 presents the 1.4 GHz radio properties of the K11 objects which reside in the surveyed 
regions of FIRST. %For completeness, Section 3.4 describes the comparison of our FIRST results to those from the NVSS survey.
Section 4.1 examines the relationship between the  1.4~GHz and X-ray properties of our BAT sample in the
 context of the black hole ``fundamental plane''.  We discuss the implications of our findings
 in Section 4.2.  A summary of our results and conclusions can be found in Section 5.
%- ref. Hickox+, Griffith\&stern(2010)  radio-selected AGNs with "red sequence" galaxies, mid-IR-selected AGNs with "blue cloud" galaxies, and X-ray-selected AGNs straddling these samples in the "green valley."   Hence, an X-ray selected sample of AGN is very interesting from the point of view of studying the crossroads of galaxy evolution.

%- Griffith\&stern(2010) find optical brightness scales with X-ray and mid-IR brightnesses, while little correlation is evident between optical and radio brightnesses. This suggests that X-ray- and mid-IR-selected AGNs have similar Eddington ratios, while radio-selected AGNs represent a different accretion mechanism with a lower and wider range of Eddington ratios.

%-Trump et al (2015)---characterisation of biases of optical emission line ratio and AGN occupation function suggest that AGNs are unlikely to be the dominant source of star formation quenching in galaxies, but instead are fueled by the same gas that drives star formation activity

\section{Sample and observations}
\subsection{The nearby Swift BAT ultra hard X-ray AGN sample}
The Burst Alert Telescope (BAT) instrument on board the Swift satellite telescope has conducted an all-sky survey in the ultra hard X-ray wavelengths, $14-195$~keV \citep{tueller10}. As the position error of the BAT observations is large \citep[$\approx 2$\arcmin; ][]{koss11a}, higher resolution X-ray observations have been obtained for nearly every source using the Swift X-ray Telescope (XRT) instrument \citep{burrows05}.  To investigate the co-evolution of AGN and their host galaxies, \citet{koss11a} selected 183 nearby AGN that reside at $z<0.05$.  Beamed AGN such as blazars have been specifically excluded from this sample.

\subsection{Complementary 1.4 GHz observations}
In this paper, we study the 1.4 GHz properties of the K11 sample primarily using the observations
 from the Faint Images of the Radio Sky at Twenty Centimeters \citep[FIRST; ][]{becker95,white97} survey\footnote{{\tt{http://sundog.stsci.edu/}}} covering $>9000$ square degrees of the Northern sky down to a 1$\sigma$ noise level of 150~$\mu$Jy~beam$^{-1}$ at 5\arcsec\ angular resolution.  We also compare the results from FIRST to the larger NRAO VLA Sky survey \citep[NVSS; ][]{condon98} which uses the VLA in a different array configuration.

The FIRST survey was selected as the primary source of the 1.4~GHz observations because the angular resolution afforded by FIRST is more consistent with  observations at shorter wavelengths and minimises confusion in the cross-identification of sources. The angular resolution of FIRST is a factor of 9 smaller than that obtained by NVSS.  While NVSS covers a larger area of sky than FIRST, NVSS is relatively shallow and has a 50\% completeness level at 2.5~mJy~beam$^{-1}$ and an angular resolution of 45\arcsec\ \citep{condon98}. 
The trade-off with greater angular resolution in interferometric observations is the likelihood of resolving out the larger angular scale diffused emission. While NVSS is unable to distinguish between multiple radio sources which are closer than 50\arcsec\ in separation,  extended NVSS sources would be resolved into several radio components by FIRST where the total flux of all components is less than a single extended Gaussian component \citep{white97}.  

 Due to these complications in comparing the flux densities between the FIRST and NVSS surveys, we use the measurements from both the FIRST and NVSS surveys in Section 4  to ensure that the results found in this paper are not biased by the technical aspects of radio interferometry.

%We compare our observations to the FIRST survey rather than the larger all-sky NRAO VLA Sky survey \citep[NVSS; ][]{condon98} because the angular resolution afforded by FIRST is more compareable to complementary observations at the shorter wavelengths. In addition, the NVSS survey is not as sensitive as the FIRST survey. On the other hand, the trade-off from the high angular resolution obtained by FIRST may be the issue of  resolving out the more diffused, larger-scale radio emission. A comparison between FIRST and NVSS will be discussed in Section 4.2.

\begin{figure}
\begin{center}
\includegraphics[scale=.75]{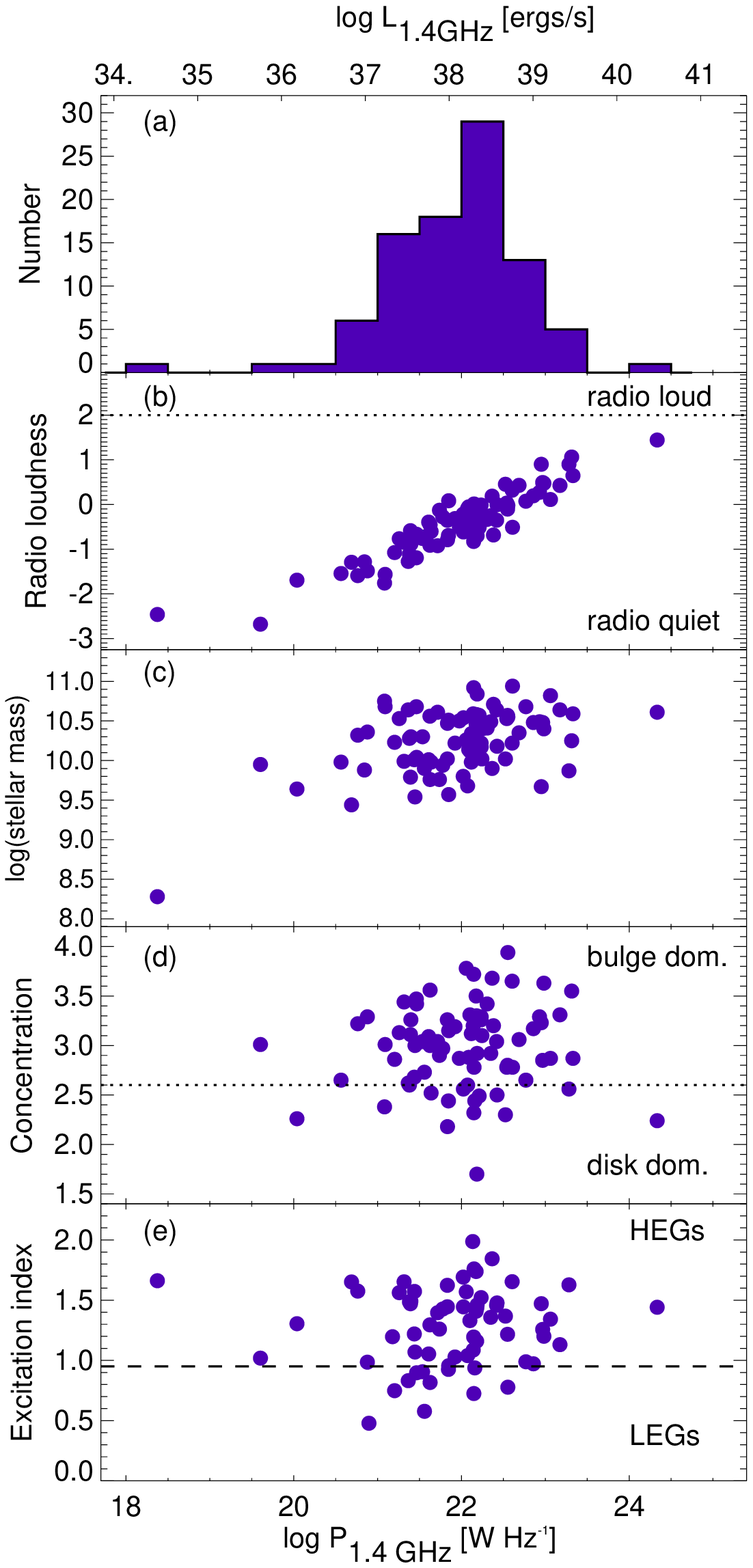} 
\end{center}
\caption{Panel (a): the 1.4 GHz radio luminosity distribution of the nearby Swift BAT AGN sample detected by the FIRST survey \citep{becker95}.   Panel (b): radio loudness \citep{ivezic02} as a function of 1.4 GHz radio luminosity. The dotted line marks the separation between radio-loud and radio-quiet sources \citep{ivezic02}.  Panel (c): the stellar masses of the host galaxies \citep{koss11a} as a function of radio luminosity. Panel (d): the structural concentration index  as a function of radio luminosity. The horizontal dotted line marks the division between bulge- and disk-dominated host galaxies \citep[e.g.\ ][]{strateva01}.  Panel (e): The excitation indices \citep{buttiglione10} as a function of the 1.4 GHz radio luminosity for the K11 sample where optical emission lines have been measured.  The dashed line separates the HEGs from the LEGs in this sample \citep{best12}.}
\label{radlum}

\end{figure}

\section{Radio 1.4 GHz properties of the Swift BAT sample}
Of the 183 BAT nearby AGN from \citet{koss11a}, we find 108 sources that are located in regions observed by the FIRST 1.4 GHz radio survey \citep{becker95}. The 1.4~GHz flux measurements for our sample are obtained from the FIRST catalogue \citep{white97}\footnote{The FIRST catalogue is available online at {\tt{http://sundog.stsci.edu/cgi-bin/searchfirst}}}.
 Table~\ref{genprop} lists the properties of these 108 BAT-selected sources. Figure~\ref{radlum}a shows the distribution of 1.4 GHz radio luminosity densities ($P_{\rm{1.4GHz}}$) of the nearby Swift BAT AGN sample. With the exception of Mrk~463,  the K11 sample detected by FIRST have 1.4 GHz luminosity densities below $10^{24}$~W~Hz$^{-1}$\footnote{A luminosity density, $P_{\rm{1.4GHz}}=10^{24}$~W~Hz$^{-1}$  corresponds to a monochromatic radio luminosity, $L_{\rm{1.4GHz}}$ of $10^{40.15}$~ergs~s$^{-1}$.}. Previous radio source population studies at 1.4 GHz have found that sources with  1.4 GHz luminosities greater than $10^{24}$~W~Hz$^{-1}$ are most likely from AGN, whereas 1.4 GHz luminosities below $10^{24}$~W~Hz$^{-1}$ consist of radio emission from both AGN and star-forming galaxies  \citep[e.g.\ ][]{rosario13,condon12,kimball11,mauch07}.  

The radio loudness of a source can be defined using the radio loudness parameter, $R$ \citep{ivezic02}:
\begin{equation}
R = 0.4 (i - t)
\end{equation}
where $i$ and $t$ represents the $i$-band and 1.4 GHz radio magnitude in the AB-magnitude system, respectively.  The AB radio magnitude $t=-2.5 log (S_{\rm{INT}}/3631 Jy)$ where $S_{\rm{INT}}$ represents the integrated flux from FIRST \citep{ivezic02}.  A radio source is considered radio-loud where $R \geq 2.0$.  The observed correlation between $R$ and $P_{\rm{1.4GHz}}$ is not surprising due to the presence of the radio flux measurements in both parameters. We find that the ultra hard X-ray selection of the K11 sample consists of radio-quiet galaxies (see Figure~\ref{radlum}b).

 For the 16 K11 sources which have not been detected by the FIRST survey,  we stacked all 16 non-detections  and found an average $P_{\rm{1.4GHz}}=1.1 \times 10^{21}$~W~Hz$^{-1}$ ($L_{\rm{1.4GHz}}=10^{37.19}$~ergs~s$^{-1}$).  We find that on average, the radio luminosity densities of our sample do not correlate with  the stellar mass nor the structural concentration of the host galaxies (panels c and d of Figure~\ref{radlum}).  A bulge-dominated galaxy is described by concentration indices greater than 2.6, while disk-dominated galaxies is described by concentration indices below 2.6 \citep{strateva01}.

% Following the definition of radio loudness from \citet{ivezic02}, we find that the ultra hard X-ray selection of the K11 sample consists of radio-quiet galaxies (see Figure~\ref{radlum}b). For the 16 K11 sources which have not been detected by the FIRST survey,  we stacked all 16 non-detections  and found an average $P_{\rm{1.4GHz}}=1.1 \times 10^{21}$~W~Hz$^{-1}$ ($L_{\rm{1.4GHz}}=10^{37.19}$~ergs~s$^{-1}$).  We find that on average, the radio luminosity densities of our sample do not correlate with  the stellar mass nor the structural concentration of the host galaxies (panels c and d of Figure~\ref{radlum}).  A bulge-dominated galaxy is described by concentration indices greater than 2.6, while disk-dominated galaxies is described by concentration indices below 2.6 \citep{strateva01}.

A hard X-ray AGN selection selects for AGN which straddle the division between high-excitation- and low-excitation- galaxies; also known as HEGs and LEGs \citep[e.g.\ ][]{hardcastle07}.
While HEGs dominate at radio luminosity densities above $P_{1.4}=10^{26}$~W~Hz$^{-1}$ ($L_{\rm{1.4GHz}}=10^{42.15}$~ergs~s$^{-1}$), both HEGS and LEGS are found at lower luminosity densities.  LEGs are also typically associated with early-type galaxies.  Conversely, HEGs can be hosted by  late-type galaxies with younger stellar populations  and have high Eddington ratios. Optical spectroscopic measurements of emission lines are obtained from \citet{koss11a} and from SDSS DR10 %for 54 of the 108 K11 sample  
(for the K11 sample with FIRST detections) in order to determine the distribution of excitation indices in our hard X-ray selected sample of nearby AGN. The excitation index ($EI$) is parameterised from six optical emission lines as follows \citep{buttiglione10,best12}:
\begin{equation}
EI = \mathrm{log} \left[ \frac{[OIII]}{[H\beta]} \right] - \frac{1}{3} \left[ \mathrm{log} \frac{[NII]}{[H\alpha]} +  \mathrm{log} \frac{[SII]}{[H\alpha]}  +  \mathrm{log} \frac{[OI]}{[H\alpha]} \right]
\end{equation}
Recent studies by \citet{best12} found that LEGs typically have $EI < 0.95$ while HEGs have $EI \geq 0.95$.  Following this definition, we find that only 17\% of our sample have excitation indices that are 
consistent with LEGs, with the remaining sample consisting of HEGs (Figure~\ref{radlum}e).

%Using optical spectroscopic measurements of emission lines  from \citet{koss11a} %for 54 of the 108 K11 sample  
% with FIRST detections, we calculate the  excitation index for each of our BAT  AGN 
%following the methods of \citet{buttiglione10} and \citet{best12}.   We find that only 17\% of our sample have excitation indices that are 
%consistent with LEGs, with the remaining sample consisting of HEGs (Figure~\ref{radlum}e). %Figure~\ref{concexcind} shows that while most of the K11 sample appear to be hosted by bulge-dominated galaxies, we do not observe a clear correlation between the excitation levels to the structural morphology of the host galaxies.

\subsection{Compact radio sources}

With the advent of the current generation of large area radio continuum surveys 
such as NVSS \citep{condon98}, FIRST \citep{becker95} and WENSS \citep{rengelink97} 
in the Northern hemisphere and SUMSS \citep{bock99,mauch03} in the Southern hemisphere; 
it is now commonly understood that the most luminous radio sources in the Universe
 are powered by AGN. However at low luminosity densities, we see an increase in 
 radio emission contribution from star-forming  galaxies to the radio luminosity 
function \citep[e.g.\ ][]{condon89,mauch07,prescott16}.

The vast majority (93.4\%) of the nearby Swift BAT AGN have been found by 
the FIRST survey to be compact radio sources in addition to being radio-quiet.
Therefore, is the radio emission originating from the star formation, or the
AGN, or a combination of both \citep{condon92,haarsma00,kimball11,condon13}? 

To further investigate the radio emission origin in this ultra hard X-ray 
selected sample, we begin by examining this sample in the context of the
radio--FIR correlation---a correlation often used to calibrate and determine
the star formation rates of galaxies \citep{vanderkruit71,helou85,condon88,yun01}.
It is often thought that this correlation is a reflection of the star 
formation process because the far-infrared emission at 60-to-250 ~$\mu$m is typically 
dominated by emission from star formation processes \citep[e.g.\ ][]{yun01,bell03}.
The infrared emission arising from AGN typically originates from the AGN torus
 that is composed of  hot dust at temperatures of $\approx 1200$--$1500$~K 
(close to the sublimation temperatures of silicate and graphite grains), which
 results in emission that peaks in the  mid-infrared (7--30~$\mu$m) wavelengths 
\citep{rodriguez06,elitzur06,schartmann05}.
 
\begin{figure}
\begin{center}
\includegraphics[scale=.45]{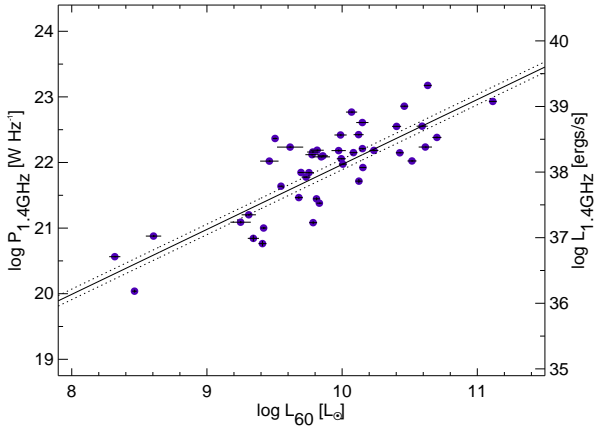}
\end{center}
\caption{The radio--FIR correlation for our sample of compact radio sources (shown
by the purple solid points).  The solid and dashed lines are not fits to our sample 
but rather, the radio--FIR correlation found by \citet{yun01}.   }
\label{radfir}
\end{figure}

We calculate the 60~$\mu$m luminosities following the calibration from 
\citet{yun01} where
log~$L_{60} (L_{\odot})= 6.014 + 2 \mathrm{log} D + log S_{60}$ and 
 the 60~$\mu$m fluxes ($S_{60}$) are from the IRAS Faint Source Catalog \citep{moshir90}.  
Figure~\ref{radfir} shows that our sample is largely consistent with the
radio--FIR correlation found by \citet{yun01} (represented by the solid black 
line).  

We would not have naively expected the FIR emission to correlate with the 1.4 GHz 
radio emission if the AGN is contributing solely to the observed radio emission 
because the  FIR emission of AGN and non-AGN galaxies are relatively similar.   
Recent results have verified indeed that it is not possible to distinguish 
between AGN and non-AGN galaxies solely from the far-infrared emission at 
60-to-250 ~$\mu$m \citep{hatziminaoglou10} and the addition of mid-infrared observations 
 is crucial for the identification of the AGN population  \citep{hatziminaoglou10,manners02}.
Therefore, to further investigate the AGN origin of the observed radio emission, we 
compare the slope of the radio--FIR correlation with the far-infrared 
spectral index between 25~$\mu$m and 60~$\mu$m (a proxy for the dust temperature within the galaxy). 

\begin{figure}
\begin{center}
\includegraphics[scale=.5]{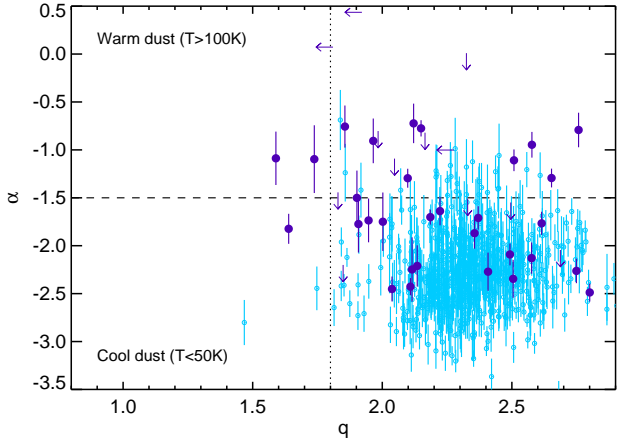}
\end{center}
\caption{Distribution of Swift BAT AGNs and star-forming galaxies using the $q$ versus $\alpha$ 
parameters where $q$ represents the FIR--radio flux ratios and $\alpha$ represents
the FIR spectral index between 60 and 25 microns. Our sample is shown by the large 
purple solid points and purple upper limits.  The small blue open circles represent the
sample of star-forming galaxies from \citet{mauch07}.  The horizontal dashed line 
separates  galaxies dominated by  `warm' ($\alpha > -1.5$) and `cool' ($\alpha<-1.5$)
 dust.  The vertical dotted line marks the $q=1.8$ value that is used to differentiate
star-forming galaxies from AGNs via the radio--FIR correlation by previous studies 
\citep[e.g.\ ][]{condon02}.
}
\label{qa} 
\end{figure}

The slope of the radio--FIR correlation is parameterised as $q$  \citep[e.g.\ ][]{helou85,condon02}:
\begin{equation}
q=log\; \Bigg[ \frac{FIR/(3.75 \times 10^{12}}{S_{\mathrm{1.4 GHz}}} \Bigg]
\end{equation}
where the FIR flux between 42.5~$\mu$m and 122.5~$\mu$m, 
$FIR$ [W~m$^{-2}]= 1.26 \times 10^{-14} [2.58 S_{60} + S_{100}]$; and $S_{60}$ and $S_{100}$ 
are the observed flux densities at 60~$\mu$m  and 100~$\mu$m (in units of Jansky).
 Following the methodology of \citet{mauch07}, we calculate the
 far-infrared spectral index as $\alpha=log(S_{25}/S_{60})/log(60/25)$
\citep[e.g.\ ][]{condon02,mauch07}.  Figure~\ref{qa} compares the distribution 
of $\alpha$ values as a function of the $q$ parameters for our BAT sample 
(purple solid points) to those of a low redshift sample of star-forming galaxies.
Similar to Figure~\ref{radfir}, the uncertainties in the 1.4 GHz measurements are 
smaller than the data points shown in Figure~\ref{qa}.

While the distribution of $q$ parameters for our sample appear similar to that of 
star-forming galaxies,  we do find a significant fraction of the nearby BAT AGN that 
appear to have warmer dust temperatures (flatter FIR spectral indices, $\alpha>-1.5$) 
as well as lower $q$ values than are typical for star-forming galaxies 
(small blue circles).  The Kolmogorov-Smirnov
test is used to compare the distribution of $q$ and $\alpha$ values that we obtained 
for the compact objects in our BAT sample to those of star forming galaxies from \citet{mauch07}.
  We find low probability values of 
$3.6\times10^{-4}$ and $1.8\times10^{-8}$ (for the $q$ and $\alpha$ values, respectively), 
meaning that the distributions for the BAT sample and that of star forming galaxies 
do not arise from the same parent distribution. Therefore, from the comparison of 
 $q$ and $\alpha$ values, it is statistically unlikely for 
the radio and FIR emission of our sample to be due solely to star formation within the host
galaxy.

So why is our sample largely consistent with the radio--FIR correlation even though the
radio emission of our sample is unlikely to have a star formation origin?  One likely
 explanation for this is that while our radio emission may be dominated by the compact
core AGN emission; the 60~$\mu$m measurements probably include emission from star
 formation  that is occuring beyond the cores of our sample galaxies, as the native 
angular resolution from IRAS is a few arcminutes in size \citep{moshir90}.  Hence, the
main conclusion from this section is that the origin of the 1.4 GHz radio emission 
is likely to include a significant AGN contribution even though our sample is 
largely consistent with the radio--FIR correlation.  We further investigate the 
AGN origin of our radio-quiet K11 sample in the context of the black hole 
fundamental plane relationship in Section 4.1

%\begin{table}
%\caption{Nearby BAT sources whose radio emission may be driven by a weak radio AGN.}
%\label{likelyagn}
%\scriptsize{
%\begin{center}
%\begin{tabular}{lccc}
%\hline
%\hline
%Galaxy & $q$ & $\alpha$ & Radio morphology\\
%\hline
%2MASXJ07595347+2323241    &   2.12 &   -0.72 & compact \\
%CGCG046-033    &   2.58 &   -0.95 & compact\\
%CGCG122-055    &   1.59  &   -1.09 & compact\\
%IC0486      & 1.64 &    -1.82 & compact\\
%MCG+01-57-016 &      &     & compact \\
%MCG+02-21-013&       2.76   & -0.79 & compact\\
%MCG+05-28-032 &      1.86 &    -0.76 & compact \\
%MCG+06-24-008 &      1.7377006  &   -1.09659 & compact \\
%Mrk198      & 1.9651762   & -0.905806 & compact \\
%Mrk268      & 2.0981055    & -1.29561 & compact \\
%Mrk290    &      &     &  compact \\
%Mrk1044     &  2.6529968  &   -1.29361 & compact \\
%Mrk1392     &  2.5080998  &   -1.10866 & compact \\
%Mrk1469     &  2.1495566  &  -0.775146 & compact \\
%Mrk704       &      &     & compact \\
%NGC1052 &      &     & extended structure \\
%NGC1142 &      &     & extended structure \\
%NGC3079      &      &     & radio jet \\
%NGC4388  &      &     & radio jet \\
%NGC5548   &      &     & radio jet \\
%\hline
%\hline
%\end{tabular}
%\end{center}}
%\end{table}
%Mrk290                                            43.42 0.02     7.9 0.02
%NGC4388                                          41.26 0.1      8.53 0.
%NGC5548                                          43.04 0.01    8.21 0.02

\begin{figure*}
\begin{center}
\includegraphics[scale=.21]{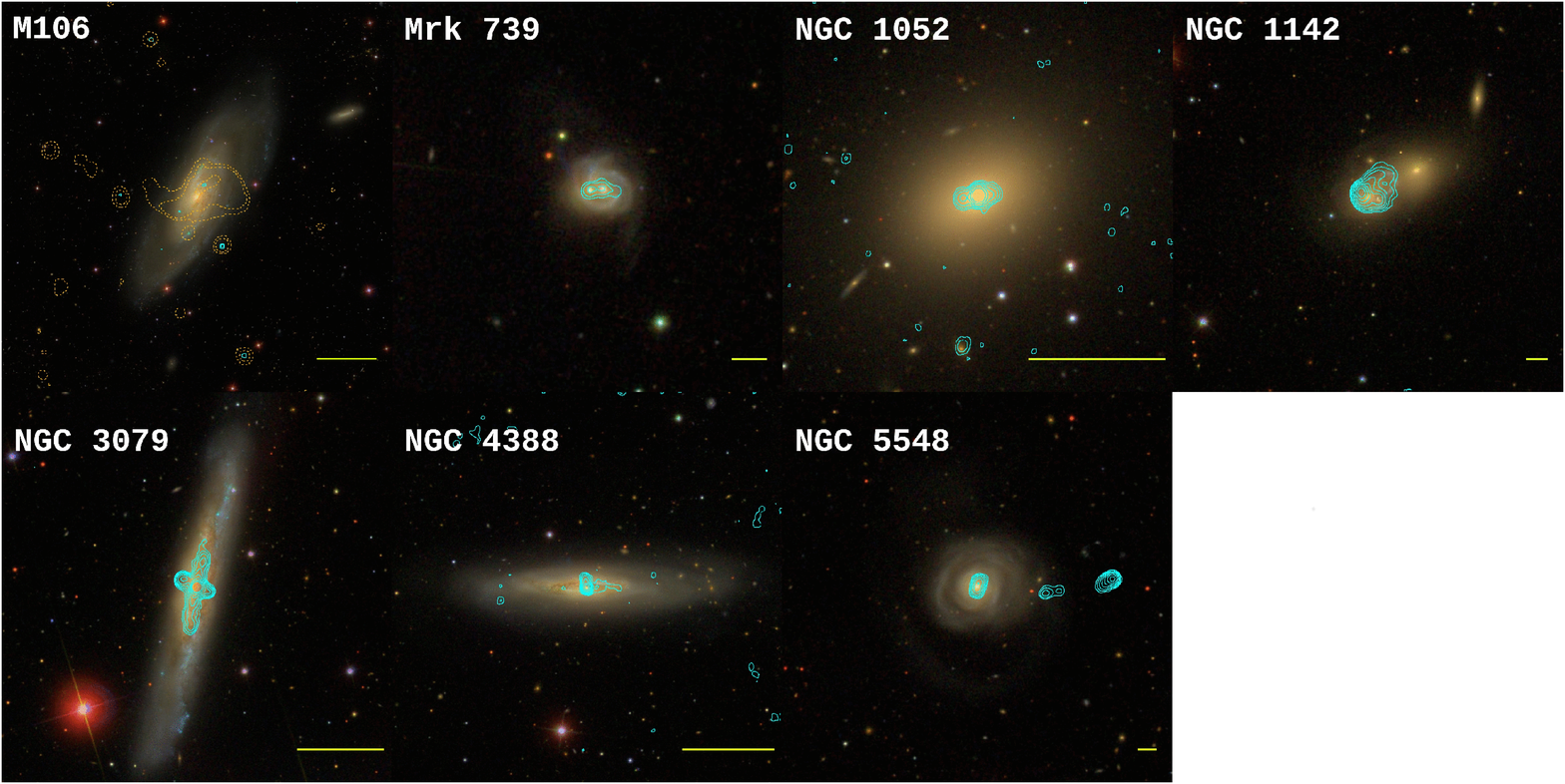}
\end{center}
\caption{Radio 1.4 GHz observations from the FIRST survey (cyan contours) and optical $ugriz$ colour maps from SDSS DR12 \citep{alam15} 
of seven Swift BAT AGN with extended radio continuum emission.   The lowest FIRST radio contour in every panel begins at  
4$\sigma$ where $1\sigma=150$~$\mu$Jy~beam$^{-1}$. Each subsequent radio contour for NGC~1052 increase by a factor of 3, while, 
the radio contours are spaced at increments of $\sqrt{3}$ for all the other galaxies. The yellow bar at the bottom-right of each
panel marks the physical scale of 10 kiloparsecs.  To better illustrate the extended radio continuum emission observed within the star-forming disk of M~106, we include the NVSS observations as orange dotted contours at the 3.5$\sigma$, 10$\sigma$ and 100$\sigma$ levels.
}
\label{extmaps} 
\end{figure*}

\subsection{Extended radio sources}

We find seven BAT sources with extended radio morphologies.  Figure~\ref{extmaps} presents the 
radio and optical maps of the K11 sample with extended radio emission.
  With the exception of M~106, the extended radio 
morphologies for the remaining six BAT sources exhibit characteristics consistent with being AGN-driven radio outflows.
The extended radio emission in NGC~1052 appears to be bipolar radio jets emanating from the central AGN.
In the cases of NGC~3079 and NGC~4388, the extended radio emission shows orthogonal radio jets as well as star formation
within the disks of the galaxies.  Due to our proximity to M~106, we observe extended radio emission due to star formation
from within the galaxy disk.

Mrk~739 is a rare binary AGN \citep{koss11-binary, teng12} and consist of Mrk~739E (the eastern core) and Mrk~739W (the western core). 
Mrk~739E has high Eddington ratio and is accreting efficiently, while, Mrk~739W only shows evidence for an AGN in hard X-ray
observations as it has very low [OIII] optical line emission.  It is possible that the extended radio emission observed to the
west of Mrk~739W is due to the inefficiently-accreting AGN in Mrk~739W.

Due to the face-on nature of NGC~5548 and the projected proximity to a pair of unrelated radio sources,
 it is unclear whether the observed extended emission within the core of the galaxy is due to star formation or the central 
supermassive black hole.  While not explicitly a merger, this galaxy does  show signs of previous interactions (as 
exemplified by the long tidal tail).

\section{Discussion}
\subsection{The radio--X-ray ``fundamental plane'' for weak radio AGN}
%X-ray selected AGN are likely to be relatively-young AGN. Relative to 
%radio-selected AGN, the X-ray selected AGNs have bluer and less evolved
%host galaxies .  %These X-ray selected AGN are likely to be growing via the  accretion of cold rather than hot gas.
We have shown in Section 3.1 that  the radio and FIR properties of our radio quiet K11 
sample is  statistically distinct from that of star-forming galaxies even though
a large number of BAT sources appear to follow the radio--FIR correlation that
 is typically populated by star-forming galaxies. The radio--X-ray ``fundamental plane'' 
correlation infers a scale invariant coupling between the accretion of matter onto a
 black hole  and the observed radio synchrotron emission, often in the form of a radio jet.
 Therefore we hypothesise that
 the K11 sample of radio quiet AGN is likely to follow the 
radio--X-ray `fundamental plane' relationship \citep[e.g.\ ][]{merloni03,falcke04,panessa15}
 if the observed  radio emission  is due to an accreting supermassive black hole 
rather than star formation from within the host galaxy.

\begin{figure*}
\begin{center}
\includegraphics[scale=1.]{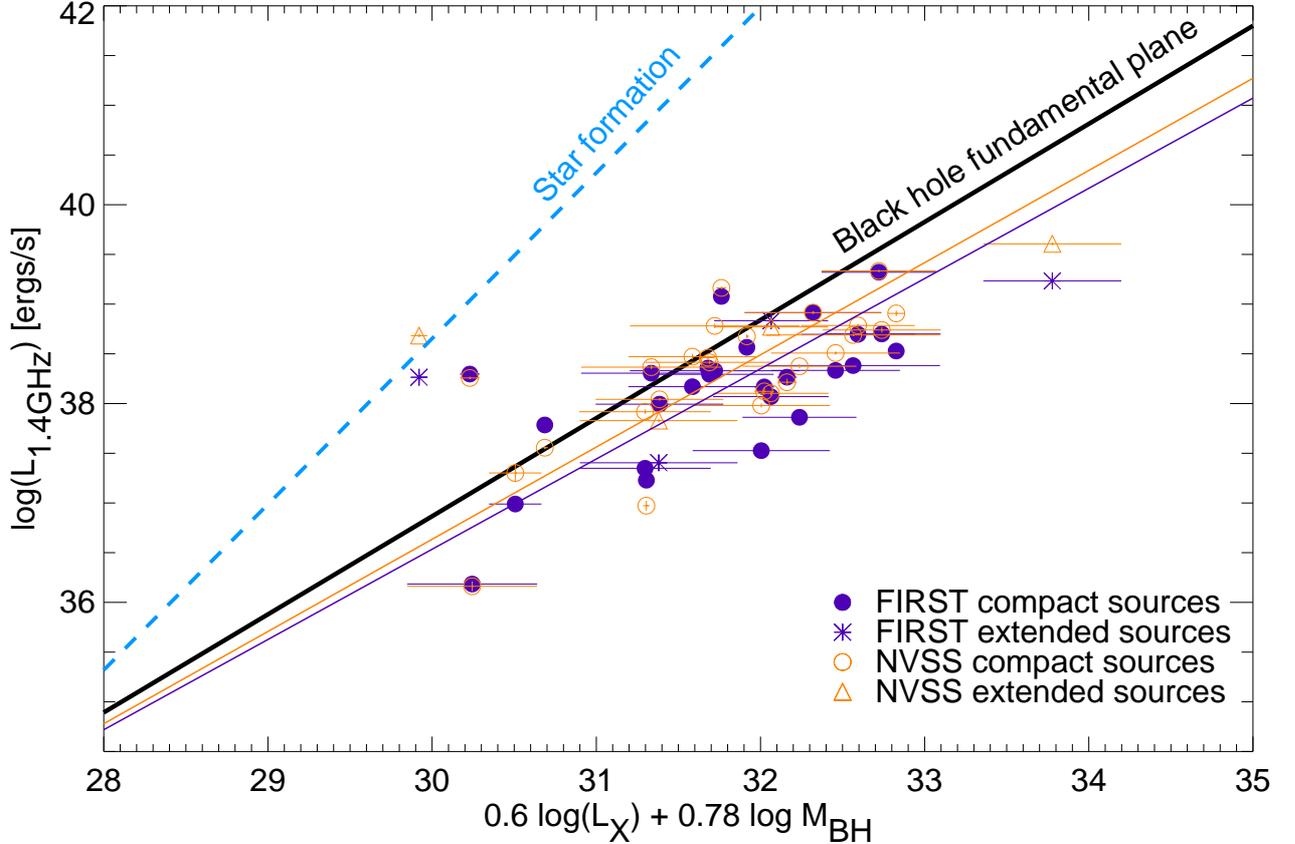}
\end{center}
\caption{The radio--X-ray ``fundamental plane'' \citep{merloni03} is represented by the black solid line. The FIRST measurements of compact and extended AGN are represented by solid purple points and purple asterisks, respectively.   The NVSS measurements of compact and extended AGN are represented by the orange open circles and orange open triangles, respectively.  Ordinary least squares bisector fits to the FIRST and NVSS observations are shown by the thin purple and thin orange lines, respectively. The thick blue dashed line shows the expected relationship if the sole origin of the observed radio and X-ray luminosities is from star formation \citep{mineo14,bell03}.  }
\label{LXLR} 
\end{figure*}

Figure~\ref{LXLR} shows the radio--X-ray `fundamental plane' relationship as 
described by \citet{merloni03} as the thick solid black line.  We represent the
FIRST measurements of the compact and extended radio sources as purple solid points
and purple asterisks, respectively.  In addition, the NVSS measurements of the 
same compact and extended radio sources are represented by the orange open circles 
and orange triangles, respectively.  We use the 1.4~GHz measurements from both 
the FIRST and NVSS surveys to rule out the potential  for the underestimation of 
 diffused radio emission from the FIRST observations.

The black hole masses for our sample are determined using the central stellar 
velocity dispersion $\sigma_\star$ and single epoch broad line H$\beta$ measurements
\citep{koss16}. The measurements assumed the the M-$\sigma$ relation from \citet{kormendy13}.   
For broad line sources, we estimate the black hole masses using an H$\beta$ fitting routine 
which includes the fitting of the Fe~II lines on the red wing as well as [O~III] ($\lambda$=4959\AA) 
as described by \citet{trakhtenbrot12}.  In addition, the 15--195~keV emission from
the Swift BAT observations were scaled by a factor of 2.67 to X-ray luminosities expected
 at 2--10~keV \citep{rigby09}.

To determine the expected correlation between the radio and X-ray emission for galaxies whose 
emission is entirely dominated by the star formation process, we assume a zero black hole mass.
  The blue dashed line in Figure~\ref{LXLR} represents the relationship that we would expect 
from a purely star-forming galaxy without a central supermassive black hole. The 
X-ray luminosity to star formation rate calibration from \citet{mineo14} and \citet{bell03}
is used to estimate the blue dashed line in Figure~\ref{LXLR}.

%The `fundamental plane' of black hole activity correlation between the 
%radio and X-ray luminosities of radio-loud AGN
% \citep[e.g.\ ][]{merloni03,falcke04,panessa15} is  observed in our sample.  
%Our current understanding of the relationship between the radio and X-ray luminosities 
%of radio-loud AGN  is that the X-ray emission is typically correlated 
%with the global radio emission and not the emission from the younger inner radio
% jet emission \citep{panessa13}. As the 1.4~GHz observations from the FIRST survey 
%may result in the underestimation of the radio emission from extended diffused AGN,  
%we use the 1.4~GHz measurements from both the FIRST and NVSS surveys for the K11
%sample for which we have previously obtained black hole mass estimates.

As can be seen from Figure~\ref{LXLR}, the radio--X-ray `fundamental plane' relationship
 provides a  useful diagnostic for differentiating between the radio emission that 
originates from black hole activity and that from star formation. This is especially  the
 case for the K11 sample where the radio emission in a majority of our sample is consistent 
with that expected from an accreting supermassive black hole. The purple and orange thin
 solid lines in Figure~\ref{LXLR} are ordinary least squares bisector fits to the
 observed measurements from the FIRST and NVSS surveys, respectively. The residual dispersions
 of these fits are 0.84 dex and 0.78 dex for the FIRST and NVSS measurements, respectively. 
While the fits to our sample (as shown by the purple and orange lines in Figure~\ref{LXLR})
 appear to be offset from the `fundamental plane' relationship as described by \citet{merloni03},
we note that this offset is not statistically significant as the relationship described 
by \citet{merloni03} lies within the uncertainties of our fits.

%Mrk290                                            43.42 0.02     7.9 0.02
%NGC4388                                          41.26 0.1      8.53 0.
%NGC5548                                          43.04 0.01    8.21 0.02

%This subsection investigates wheter the 1.4 GHz radio luminosities obtained from the FIRST survey is representative of the total radio emission output from these radio-quiet sources, we compare the 1.4 GHz radio luminosities of our weak radio AGN sources (Table 1) from the FIRST survey to observations at the same frequency by the NVSS survey which has a larger angular resolution, lower flux sensitivity but  more likely to detect more diffused radio emission which may be resolved out by FIRST.  

The two largest outliers to the `fundamental plane' are NGC~3079 and Mrk~766.
NGC~3079 has an extended radio morphology which includes radio emission from 
its star forming disk for which we have not subtracted.  Mrk~766 is a narrow
 line Seyfert 1 AGN hosted by a  nearby star forming spiral galaxy 
where a 60 parsec extended radio outflow have been
previously observed at 3.6~cm by \citet{ulvestad95}. Similar to NGC~3079, the 
 synchrotron emission from star formation within Mrk~766 is likely to dominate
the observed 1.4 GHz emission even though a radio jet is present.
 In addition to greater uncertainties in
the measurements of diffused radio emission \citep{white97}, it is also known that 
 the connection to the radio jet 
emission mechanism may depend on timescale and observed radio frequency \citep{burlon13}. 
%Therefore it is possible that  offsets between the fits to our data and the 
%original radio--X-ray fundamental plane \citep{merloni03} is due to a time lag 
%between the observed X-ray AGN and the onset of a radio AGN.

 Our current understanding of the relationship between the radio and X-ray luminosities 
of radio-loud AGN  is that the X-ray emission is typically correlated 
with the global radio emission and not the emission from the younger
parsec-scale inner radio jet emission \citep{panessa13,burlon13,panessa15}. 
 In addition, the radio--X-ray comparison by \citet{burlon13} at 20~GHz found a
 significantly different slope to previous studies which investigated the
 radio emission at lower frequencies \citep[e.g.\ ][]{merloni03,panessa15}. 
 However, the results from \citet{burlon13} are likely to be hampered by
 distance effects and the inclusion of beamed blazars in their sample.

While the accretion disk--radio jet coupling has often been thought to be the
 physical interpretation for the radio--X-ray `fundamental plane' \citep[e.g.\ ][]{doi11}, recent 
 studies such as \citet{behar15} and \citet{kharb15} have proposed alternative
non-jet origins for the observed  radio emission from radio quiet AGN 
which may also account for any observed offset in the `fundamental plane' 
between our K11 sample and those 
 of nearby Seyferts from \citet{merloni03}. Unlike radio loud AGN, the radio
emission from radio quiet AGN can originate from the accretion disk 
corona \citep{behar15,kharb15}. The corona radio emission are neither collimated 
nor relativistic and are analogous to stellar coronal mass ejections \citep{bastian98,bastian07}. 
If this is the true explanation for the observed radio emission in radio quiet AGN, 
 the corollary is a lower radio-to-X-ray luminosity ratio than that
observed in radio loud AGN \citep{behar15}. However, as we have not observed
a statistically significant decrease in the radio-to-X-ray luminosity ratio 
for our sample, relative to that found from \citet{merloni03}, we are not
able to confirm the possibility of the non-jet origin for the radio emission
observed in our sample.

%In addition to the issue of timescale, 
%the non-jet origin for the source of the observed radio emission can explain the  
%lower $L_{\rm{1.4GHz}}$/$L_{\rm{X}}$  observed for the majority of our 
%radio quiet K11 sample relative to the `fundamental plane' found by \citet{merloni03}.

In summary, we have found that the majority of our radio quiet K11 sample is consistent with
the general black hole `fundamental plane' even though our sample appears to 
be consistent with the radio--FIR correlation that we expect from star-forming
 galaxies.  While it is possible for the observed radio emission 
to be dominated by star formation from the host galaxy, our `fundamental plane'
 results indicate that star formation alone is not able to explain the observed 
radio and X-ray emission from these radio quiet AGN. Our results are consistent
with other recent studies that make the argument that the radio emission of radio quiet 
quasars are AGN-related \citep{zakamska16,barger15,white15}. 
Related to these results are those from previous studies by \citet{ho01,nagar02} 
who found significant detections of radio sources with  both compact and jet 
morphologies in radio surveys of optically-selected low luminosity AGN.
This is in contrast to
to other results that find that the radio emission from radio quiet galaxies
are more consistent with that of star-forming galaxies \citep[e.g.\ ][]{bonzini15}.

%--->compare the radio luminosity with the Eddington ratio in the sample
%that is most likely to host an AGN.

\subsection{Implications}
%We find that an X-ray selection of nearby AGN that mainly represents Green Valley bulge-dominated
%galaxies regardless of their excitation levels or accretion efficiencies.  
The similarity of the K11 radio and FIR properties
 to the radio--FIR correlation for star forming galaxies is consistent with previous
studies such as \citet{moric10} who found that a significant number of  Seyfert galaxies also
follow the radio--FIR correlation.  Therefore, caution should be exercised  when using the
 radio--FIR correlation to differentiate between galaxies dominated by AGN versus star formation 
 as well as to infer accurate star formation rates in galaxies that host an actively
accreting, albeit radio quiet, supermassive black hole.

While some studies suggest that the  accretion mechanism for radio-selected AGN is
 different to that of optical- or X-ray selected AGN samples \citep[e.g.\ ][]{griffith10}, 
we argue that the `fundamental plane' correlation between the radio and X-ray luminosities 
\citep{merloni03,falcke04}  support the idea that the radio and X-ray properties 
are  likely to be linked by the accretion mechanism.  We hypothesise that the difference is rather 
in the origin of the observed radio emission.  The non-jet alternative origin for the observed radio
emission, that is,  the accretion disk corona origin for the observed radio emission 
\citep{behar15,kharb15} can explain any observed offset in our sample relative
to the `fundamental plane' from \citet{merloni03} as well as  the consistency between the K11 sample
and the radio--FIR correlation observed in star forming galaxies.
A corona origin for the observed radio emission is likely to be correlated to the 
FIR luminosities and dust temperatures. In fact, a recent study by \citet{smith14} 
 has found a temperature dependence in the radio--FIR correlation in the Herschel-ATLAS
 survey, whereby 250~$\mu$m sources with warm dust 
spectral energy distributions have larger 1.4 GHz luminosities than  cooler sources.

%However, the connection to the radio jet 
%emission mechanism may depend on timescale and observed radio frequency \citep{burlon13}. 

While many simulations have used the AGN radio-mode as a regulator of star formation within 
galaxies \citep[e.g.\ ][]{croton06}, a recent study by \citet{kaviraj14} have found that
the onset of the radio AGN is relatively late (after a few dynamical timescales) into the 
associated starburst phases. As such, the radio mode of AGN is unlikely to affect the star 
formation history of the galaxy. On the other hand, given the ubiquity of accreting supermassive
black holes, our results  show that radio quiet AGN are still injecting additional radio 
feedback into its surrounding medium and  provide observational evidence for the radio maintenance 
mode that have long been proposed by simulations.
An important implication of our result is that the observed differences between
optically-selected and radio-selected samples of AGN cannot be simply attributed
to the accretion efficiencies.

\section{Summary and conclusions}

We find that the ultra hard X-ray selection of AGN from \citet{koss11a} 
have radio properties that appear to follow the radio--FIR correlation
 that we expect for star-forming galaxies. However, further analysis
of our K11 sample as a population and as individual sources show that 
 the observed radio emission is more consistent with with the processes 
associated with accreting supermassive black holes.

We show that our sample of BAT sources  follows the radio--X-ray `fundamental plane'
 correlation---an indication that  the observed radio emission 
is related to the accretion processes traced by the observed X-ray emission.
%In addition, we hypothesise that an accretion disk corona origin for the observed
% radio emission in radio quiet AGN can explian the lower $L_{\rm{1.4GHz}}$/$L_{\rm{X}}$ 
%fraction as well as the observed consistency of our sample with the 
%radio--FIR correlation.  
Consistent with recent studies of radio quiet AGN \citep[e.g.\ ][]{zakamska16,barger15,white15}, 
we conclude that radio quiet AGN includes a non-negligible component
of AGN-driven radio emission and that the radio emission from galaxies hosting 
radio quiet AGN does not 
solely consist of radio emission from star formation even if the radio and FIR
emission is mostly consistent with the radio--FIR correlation.

\vspace{1cm}

\chapter{\bf{Acknowledgments.}}
MJK acknowledges support from the Swiss National Science Foundation and Ambizione fellowship grant PZ00P2\_154799/1.  KS is supported by SNF Grant PP00P2 138979/1. ADK acknowledges financial support from the Australian Research Council Centre of Excellence for All-sky Astrophysics (CAASTRO), through project number CE110001020.  CR acknowledges financial support from the CONICYT-Chile grants ``EMBIGGEN" Anillo ACT1101 (CR), FONDECYT 1141218 (CR), and Basal-CATA PFB--06/2007.  We thank the referee for their suggestions that improved this paper.
This research has made use of the NASA/IPAC Extragalactic Database
(NED) and the NASA/IPAC Infrared Science Archive (IRSA),
 which is operated by the Jet Propulsion Laboratory, California Institute of 
Technology, under contract with the National Aeronautics and Space Administration.

\bibliographystyle{mnras} 
\bibliography{mn-jour,paperef}

\clearpage
\onecolumn

\begin{center}
\begin{longtable}{lcccccc}
\caption{The properties of the 108 nearby Swift BAT AGN sample that are located in regions observed by the FIRST 1.4 GHz radio survey.}\label{genprop} \\
%\begin{tabular}{lcccccc}
\hline \hline \multicolumn{1}{l}{Galaxy} & \multicolumn{1}{c}{RA} &  \multicolumn{1}{c}{Declination} &  \multicolumn{1}{c}{Redshift} &  \multicolumn{1}{c}{Distance} &  \multicolumn{1}{c}{log ($M_{\star}$)} &  \multicolumn{1}{c}{$S_{1.4}$}\\
\hline
\endfirsthead

\multicolumn{7}{c}%
{{\tablename \thetable{} -- continued from previous page}}\\
\hline \multicolumn{1}{l}{Galaxy} & \multicolumn{1}{c}{RA} &  \multicolumn{1}{c}{Declination} &  \multicolumn{1}{c}{Redshift} &  \multicolumn{1}{c}{Distance} &  \multicolumn{1}{c}{log ($M_{\star}$)} &  \multicolumn{1}{c}{$S_{1.4}$}\\
\hline
\endhead

\hline \multicolumn{7}{r}{Continued on next page}\\ \hline
\endfoot

\hline \hline
\endlastfoot

%  Galaxy & RA (J2000) &Declination (J2000) & Redshift & Distance & log($M_{\star}$)& $S_{1.4}$  \\ 
%(1) & (2) & (3) & (4) &(5) & (6) & (7)\\
%\hline 
2MASX J01064523+0638015 &      01:06:45.2   &    06:38:01.6  &0.041  &178.7 & 10.4 & 1.8    \\ 
MCG -01-05-047          &      01:52:49.0    &   -03:26:48.6   &0.016 &72.2 &     10.5 & 15.1  \\ 
NGC 788                 &      02:01:06.5    &   -06:48:57.0   &0.013 &58.0 &     10.68 & 3.0  \\
Mrk 1018                &      02:06:16.0   &   -00:17:29.3 &0.042 &186.1 &     10.92 &3.4  \\
ARP 318                 &      02:09:20.9   &   -10:07:59.2 &0.013 &56.3 &     ---   &2.1  \\
Mrk 590                 &      02:14:33.6   &   -00:46:0.3  &0.026 &113.7 &     10.84 &9.9  \\
Mrk 1044                &      02:30:05.5  &   -08:59:53.6  &0.016 &70.4 &     9.88  &1.2  \\
NGC 985                  &     02:34:37.9    &  -08:47:17.0 &0.043 &187.7 &    10.71 &5.7   \\
NGC 1052                 &     02:41:04.8  &  -08:15:20.7   &0.005 &19.6 &    10.35 &1053.0   \\
Mrk 595                  &     02:41:34.9    &   07:11:14.0 &0.027 &116.4 &   10.3 & ---\\
MCG -02-08-014           &     02:52:23.4    &  -08:30:37.7 &0.017 &71.7 &    10.01 &6.6   \\
NGC 1142                 &     02:55:12.3    &  -00:11:01.8 &0.029 &124.6 &    10.93 &$<0.3$   \\
NGC 1194                 &     03:03:49.1    &  -01:06:13.0  &0.013 &56.9 &    10.32 &1.5   \\
2MASX J03305218+0538253  &     03:30:52.2    &   05:38:25.3 &0.046 &201.3 &    10.02 &3.6   \\
Mrk 79			 &     07:42:32.9  &    49:48:35.0 &0.022 &95.4 &  10.57  &14.9    \\
UGC 03995                &     07:44:09.1   &   29:14:50.7  &0.016 &68.0 &10.57 & ---\\
Mrk 10			 &     07:47:29.0     &  60:56:00.9 &0.029 &126.4 & 10.73 & ---\\
2MASX J07595347+2323241  &     07:59:53.5    &   23:23:24.2  &0.030 &128.0 &    10.57 &18.0   \\
IC 0486                  &     08:00:21.0    &   26:36:48.3  &0.027 &117.5 &    10.61 &3.1   \\
MCG +02-21-013           &     08:04:46.4    &   10:46:36.3  &0.034 &149.3 &    10.68 &21.9    \\
CGCG 031-072             &     08:14:25.3    &   04:20:32.4  &0.033& 143.2 &    10.27 & ---\\
MCG +11-11-032           &     08:55:12.6    &   64:23:45.2 & 0.036& 156.3 &    10.33 & ---\\
Mrk 18                   &     09:01:58.4    &   60:09:06.2 &0.011 &47.2 &    9.57  &26.3   \\
2MASX J09043699+5536025  &     09:04:37.0    &   55:36:02.7 &0.037 &161.4 &    9.76 &1.4    \\
2MASX J09112999+4528060  &     09:11:30.0    &   45:28:06.0 &0.026 &115.5 &    9.76 &3.4    \\
Mrk 704                  &     09:18:26.0    &   16:18:19.7  &0.029 &126.3 &    10.34 &6.9   \\
SBS 0915+556             &     09:19:13.2    &   55:27:55.2  &0.049 &213.0 &    ---  &2.5  \\
IC 2461                  &     09:19:58.0    &   37:11:27.8  &0.008 &72.9 &    9.54 &4.4        \\
MCG +04-22-042           &     09:23:43.0    &   22:54:32.4  &0.033 &141.5 &    10.49 &9.3   \\
Mrk 110                  &     09:25:12.9    &   52:17:10.5  &0.035 &153.6 &    9.9  &8.2    \\
Mrk 705                  &     09:26:03.2   &   12:44:04.1  &0.028 &120.8 &    10.18 &8.5   \\
NGC 2885                 &     09:27:18.5    &   23:01:12.1  &0.026 &112.0 &    10.56 &2.8   \\
CGCG 312-012             &     09:29:37.9   &   62:32:38.3  &0.026 &110.3 &    10.3 &2.4    \\
CGCG 122-055             &     09:42:04.8   &   23:41:06.6  &0.021 &91.8 &    9.94  &5.9   \\
IC 2515                  &     09:54:39.4    &   37:24:30.8  &0.019 &82.7 &    10.04 &3.6   \\
NGC 3079                 &     10:01:57.8    &   55:40:47.2  &0.004 &19.3 &    9.98  &293.0   \\
NGC 3227                 &     10:23:30.6      &  19:51:54.3	  &0.003 &20.9 &   9.98   &82.8   \\
MCG +06-24-008           &     10:44:49.0    &   38:10:52.5  &0.026 &111.5 &    10.23& 9.7    \\
UGC 05881                &     10:46:42.5    &   25:55:54.0  &0.021 &88.1 &    10.22 &9.0   \\
Mrk 417                  &     10:49:30.9    &   22:57:52.4  &0.033 &142.1 &   10.2 & ---\\
Mrk 728                  &     11:01:01.8   &   11:02:48.8  &0.030 &154.7 &    10.09 &2.9   \\
Mrk 732                  &     11:13:49.7    &   09:35:10.7  &0.029 &126.3 &    10.48 &4.7   \\
ARP 151                  &     11:25:36.2    &   54:22:57.3  &0.021 &90.5 &     9.7 &3.0    \\
NGC 3718                 &     11:32:34.9    &   53:04:04.2 &0.003 &17.0 &    9.98 &10.6    \\
UGC 06527                &     11:32:40.2    &   52:57:01.1 &0.026 &113.9 &    10.22 &25.8   \\
IC 2921                  &     11:32:49.3    &   10:17:47.4  &0.044 &191.0 &    ---  & ---\\
Mrk 739E                 &     11:36:29.3    &   21:35:45.0  &0.030 &128.4 &    10.47 &3.4   \\
NGC 3786                 &     11:39:42.5    &   31:54:33.7  &0.009 &41.6 &    10.01 &13.2   \\
KUG 1141+371             &     11:44:29.9    &   36:53:08.6 &0.038 &165.5 &   10.19 & ---\\ 
UGC 06732                &     11:45:33.2    &   58:58:40.0  &0.010 &42.3 &    ---   &4.7   \\
MCG +05-28-032           &     11:48:46.0    &   29:38:28.6  &0.023 &99.0 &    10.14  &10.4  \\
IC 0751                  &     11:58:52.6    &   42:34:13.7  &0.031 &134.1 &    10.48 &43.0   \\
2MASX J12005792+0648226  &     12:00:57.9    &   06:48:22.7  &0.036 &156.1 &    10.58 &5.2   \\
Mrk 1310                 &     12:01:14.4    &  -03:40:41.0  &0.019 &86.1 &    ---   &3.1   \\
NGC 4051                 &     12:03:09.7   &   44:31:52.5  &0.002 &14.5 &     9.44 &19.3   \\
ARK 347                  &     12:04:29.6    &   20:18:58.1  &0.023 &96.9 &    10.3  &2.2   \\
UGC 07064                &     12:04:43.4    &   31:10:38.0  &0.025 &107.6 &    10.54 &7.6   \\
NGC 4102                 &     12:06:23.1    &   52:42:39.4  &0.003 &21.0 &     9.68 &223.4   \\
Mrk 198                  &     12:09:14.1    &   47:03:30.3  &0.025 &105.8 &    10.12&9.4    \\
NGC 4138                 &     12:09:29.8    &   43:41:07.2 &0.003 & 15.6 & 9.61 & ---\\
KUG 1208+386             &     12:10:44.3    &   38:20:10.2  &0.023 &98.0 &    10.02 &5.9   \\
Mrk 1469                 &     12:16:07.1   &   50:49:30.1  &0.031 &134.1 &    10.22 &7.9   \\
NGC 4235                 &     12:17:09.9   &   07:11:30.0  &0.008 &35.1 &    10.36 &5.1   \\
Mrk 766                  &     12:18:26.5    &   29:48:46.2  &0.013 &54.0 &    10.02 &40.3   \\
M 106                    &     12:18:57.6    &   47:18:13.4  &0.002 &7.5 &    9.95  &93.0   \\
Mrk 50                   &     12:23:24.1    &   02:40:44.4  &0.023 &99.7 &   9.9 & ---\\  
NGC 4388                 &     12:25:46.8    &   12:39:43.5  &0.008 &18.3 &    10.53 & 45.0    \\
NGC 4395                 &     12:25:48.9    &   33:32:48.3  &0.001 &4.1 &    8.28 & 1.2    \\
NGC 4593                 &     12:39:39.5     &  -05:20:39.2    &0.009 &44.0 &    10.75 &5.2   \\
NGC 4686                 &     12:46:39.8    &   54:32:03.1 &0.017 &71.7 &    10.68 &4.7   \\
SBS 1301+540             &     13:03:59.4    &   53:47:30.1  &0.030 &129.2 &    9.79 &1.2    \\
NGC 4992                 &     13:09:05.6   &   11:38:02.7 &0.025 &108.5 &    10.64 &1.6   \\
UGC 08327                &     13:15:17.3    &   44:24:25.9  &0.037 &158.8 &    10.55 &11.8   \\
NGC 5106                 &     13:20:59.6    &   08:58:42.2  &0.032 &138.3 &    10.59 &93.0   \\
NGC 5231                 &     13:35:48.3    &   02:59:55.6  &0.022 &93.5 &    10.51 &6.7    \\
NGC 5252                 &     13:38:15.9    &   04:32:33.0  &0.022 &95.5 &    10.59 &12.6   \\
Mrk 268                  &     13:41:11.2    &   30:22:41.1  &0.040 &173.6 &    10.64 &41.3   \\
NGC 5273                 &     13:42:08.4   &   35:39:15.3  &0.004 &17.7 &    9.64 & 2.9    \\
CGCG 102-048             &     13:44:15.7    &   19:33:59.7  &0.027 &116.0 &    10.31 &2.2   \\
NGC 5290                 &     13:45:19.1    &   41:42:44.4  &0.009 &35.7 &    10.23 & 10.4   \\
UM 614                   &     13:49:52.8    &   02:04:45.0  &0.033 &142.3 &    9.99 & 0.9    \\
Mrk 464                  &     13:55:53.5    &   38:34:28.7  &0.051 &222.8 &    9.67 &15.0    \\
Mrk 463                  &     13:56:02.8   &   18:22:17.3  &0.050 &221.0 &    10.61 &367.8   \\
CGCG 046-033             &     14:00:18.4    &   05:02:42.2  &0.034 &148.7 &    10.18 &10.0   \\
NGC 5506                 &     14:13:14.9    &  -03:12:27.2  &0.006 &28.7 &    10.02 &337.9   \\
NGC 5548                 &     14:17:59.5    &   25:08:12.4  &0.017 &71.4 &    10.46 & 24.4    \\
NGC 5610                 &     14:24:22.9    &   24:36:51.4  &0.017 &72.3 &    10.28 &22.4   \\
NGC 5674                 &     14:33:52.3    &   05:27:29.8  &0.025 &107.3 &    10.57 &11.0   \\
NGC 5683                 &     14:34:52.5    &   48:39:42.9  &0.036 &157.3 &    10.22 & ---\\
Mrk 817                  &     14:36:22.1     &  58:47:38.9   &0.031 &135.2 &   10.41 & 9.2   \\
Mrk 477                  &     14:40:38.1    &   53:30:15.0  &0.038 &164.0 &    9.87  &59.2   \\
2MASX J14530794+2554327  &     14:53:07.9   &   25:54:32.8  &0.049 &213.0 &     --- & ---\\
Mrk 841                  &     15:04:01.2   &   10:26:16.4  &0.036 &158.2 &    9.97 & ---\\
Mrk 1392                 &     15:05:56.6    &   03:42:26.2  &0.036 &157.3 &    10.64 &8.8   \\
NGC 5899                 &     15:15:03.3   &   42:02:59.5  &0.009 &43.9 &    10.28  &10.4  \\
NGC 5940                 &     15:31:18.1    &   07:27:27.7  &0.034 &147.4 &   10.47 & ---\\
Mrk 290                  &     15:35:52.4    &   57:54:09.5 &0.030 &128.1 &    9.8   &5.3   \\
2MASX J16174158+0607100  &     16:17:41.6    &   06:07:10.0  &0.038 &164.8 &   10.14 & ---\\
CGCG 198-020             &     17:12:28.4    &   35:53:02.3 &0.026 &112.0 &   10.38 & ---\\
ARP 102B                 &     17:19:14.5     &  48:58:49.4    &0.024 &104.0 &    10.25 &158.7   \\
CGCG 300-062             &     17:43:17.3    &   62:50:20.8  &0.033 &143.0 &     9.9  & 1.5   \\
Mrk 520                  &     22:00:41.4    &   10:33:08.0 &0.027 &114.7 &    10.4  &60.8   \\
MCG +01-57-016           &     22:40:17.1    &   08:03:13.4  &0.025 &107.5 &    10.26 &8.2   \\
NGC 7469                 &     23:03:15.7     &  +08:52:25.3    & 0.016& 69.8&    10.49 &145.7   \\
Mrk 926                  &     23:04:43.5    &  -08:41:08.5 & 0.047&206.5 &    10.82 &22.5    \\
NGC 7603                 &     23:18:56.6    &   00:14:37.6  &0.029 &126.6 &    10.94 &21.1   \\
NGC 7679                 &     23:28:46.7     &  +03:30:41.0    &0.017 &73.4 &    10.17  &26.6  \\
NGC 7682                 &     23:29:03.9     &  +03:32:00.0    &0.017 &73.1 &    10.53  &54.0  \\
%\hline                  
%\hline                  
%\end{tabular}
\end{longtable}
Col.\ (1): Object identification.  Col.\ (2): Galaxy centre's right ascension (J2000). Col.\ (3): Galaxy center's declination (J2000).  Col.\ (4): Redshift. Col.\ (5): Distance in Megaparsecs. Col.\ (6): log stellar mass (M$_{\odot}$).  Col.\ (7): Total 1.4 GHz flux in mJy from FIRST.
\end{center}
%Col.\ (1): Object identification.  Col.\ (2): Galaxy centre's right ascension (J2000). Col.\ (3): Galaxy center's declination (J2000).  Col.\ (4): Redshift. Col.\ (5): Distance in Megaparsecs. Col.\ (6): log stellar mass (M$_{\odot}$).  Col.\ (7): Total 1.4 GHz flux in mJy from FIRST.
%\end{longtable} 
\twocolumn
\clearpage

\end{document}